\documentclass[11pt]{article}

\usepackage{amsfonts,amssymb,graphicx}
\usepackage{amsfonts}
\usepackage{amsmath}
\usepackage{amstext}
\usepackage{amsthm}
\usepackage{xspace}

\newcommand{\N}{\mathbb N}
\newcommand{\C}{\mathbb C}

\newcommand{\R}{\mathbb R}
\newcommand{\Q}{\mathbb Q}

\newcommand{\set}[1]{\left\{#1\right\}}

\newcommand{\virg}[1]{``#1"}
\newcommand{\vsni}{\vskip 0.2cm}
\newcommand{\comb}[2]{\left( \begin{array}{c} #1 \\ #2 \end{array} \right)}

\newcommand\BB{{\cal B}}

\newcommand\EE{{\cal E}}
\newcommand\FF{{\cal F}}

\newcommand\HH{{\cal H}}

\newcommand\JJ{{\cal J}}

\newcommand\PP{{\cal P}}

\newtheorem {theorem}{Theorem}[section]

\newtheorem {lemma}[theorem]{Lemma}
\newtheorem {proposition}[theorem]{Proposition}
\newtheorem {corollary}[theorem]{Corollary}

\newtheorem {example}[theorem]{Example}

\newtheorem {remark}[theorem]{Remark}

\newtheorem {definition}[theorem]{Definition}
\theoremstyle{definition}
\numberwithin{equation}{section}

\renewcommand {\Re}{{\rm Re}}

\def\be{\begin{equation}}
\def\ee{\end{equation}}
\newcommand{\Leq}[1]{\label{#1}\end{equation}}
\newcommand{\beqn}{\begin{eqnarray}}
\newcommand{\eeqn}{\end{eqnarray}}
\newcommand{\beqno}{\begin{eqnarray*}}
\newcommand{\eeqno}{\end{eqnarray*}}

\begin{document}

\title{Spectral analysis of transfer operators associated to Farey fractions}

\author{Claudio Bonanno
\thanks{Dipartimento di Matematica Applicata,
Universit\`a di Pisa, via F. Buonarroti 1/c, I-56127 Pisa, Italy,
email: $<$bonanno@mail.dm.unipi.it$>$} \and Sandro Graffi
\thanks{Dipartimento di Matematica, Universit\`a di Bologna,
Piazza di Porta S. Donato 5, I-40127 Bologna, Italy, e-mail:
$<$graffi@dm.unibo.it$>$} \and Stefano Isola
\thanks{Dipartimento di Matematica e Informatica, Universit\`a
di Camerino, via Madonna delle Carceri, I-62032 Camerino, Italy.
e-mail: $<$stefano.isola@unicam.it$>$}}

\maketitle

\begin{abstract}
\noindent
The spectrum of a one-parameter family of signed transfer operators associated to the Farey map is studied in
detail. We show that when acting on a suitable Hilbert space of analytic functions they are self-adjoint and exhibit absolutely
continuous spectrum and no non-zero point spectrum. Polynomial eigenfunctions when the parameter is a negative half-integer are
also discussed.
\end{abstract}


\vskip 0.5cm
\noindent
{\sc Riassunto:} {\sl Analisi spettrale di operatori di trasferimento associati alle frazioni di Farey.}

\noindent
Presentiamo uno studio dettagliato dello spettro di una famiglia ad un parametro di operatori di trasferimento segnati
associati  alla trasformazione di Farey dell'intervallo unitario in s\'e.
Se fatti agire su un opportuno spazio di Hilbert di funzioni analitiche essi risultano autoaggiunti e con spettro
assolutamente continuo (ad eccezione dell'autovalore nullo).  
Diamo altres\`i una a classificazione completa delle autofunzioni polinomiali quando il parametro \`e un semintero negativo.  

\vskip 0.5cm
\noindent
{\sc Scientific Chapter:} Mathematical Physics

\section{Preliminaires and statement of the main results} \label{map}

\noindent
Let $F:[0,1]\to [0,1]$ be  the {\sl Farey map} defined by
\begin{equation} \label{farey}
F(x)=\left\{
\begin{array}{ll}
\frac{x}{1-x} & \mbox{if }\ 0\le x\le \frac{1}{2}\\[0.3cm]
\frac{1-x}{x} & \mbox{if }\ \frac{1}{2} \le x \le 1
\end{array} \right.
\end{equation}
Its name can be related to the following observation. If we
expand  $x\in [0,1]$ in continued fraction, i.e.
$$x = \frac{1}{a_1 + \displaystyle  \frac{1}{a_2 + \displaystyle \frac{1}{a_3 +\cdots}}} \equiv
[a_1,a_2,a_3,\dots]$$
then
\begin{equation} \label{farey-cont-frac}
x=[a_1,a_2,a_3,\dots] \longmapsto F(x)=[a_1-1,a_2,a_3,\dots]
\end{equation}
with $[0,a_2,a_3,\dots] \equiv [a_2,a_3,\dots]$. Differently said,
let $\FF_n$ be the ascending sequence of irreducible fractions
between $0$ and $1$ constructed inductively in the following way:
set first $\FF_1=( \frac 0 1, \frac 1 1)$, then $\FF_n$ is
obtained from $\FF_{n-1}$ by inserting among each pair of
neighbours $\frac {a'}{ b'}$ and $\frac{a''}{b''}$ in $\FF_{n-1}$ their
Farey sum $\frac{a}{b}:=\frac{a'+a''}{b'+b''}$. Thus
$$\FF_2=\left({\textstyle \frac 0 1, \frac 1 2, \frac 1 1} \right)
\quad \FF_3=\left(\textstyle{ \frac 0 1, \frac 1 3, \frac 1 2,
\frac 2 3, \frac 1 1} \right) \quad \FF_4=\left(\textstyle{ \frac
0 1, \frac 1 4, \frac 1 3, \frac 2 5, \frac 1 2, \frac 3 5, \frac
2 3, \frac 3 4, \frac 1 1} \right)$$ and so on. The elements of
$\FF_{n}$ are called \emph{Farey fractions}. It is easy to verify
that the set of pre-images $\cup_{k=0}^{n} F^{-k}\set{0}$
coincides with $\FF_n$ for all $n\ge 1$. This implies that
$\cup_{k=0}^\infty F^{-k}\set{0}=\Q \cap [0,1]$. These two observations are related by the fact that
a rational number $\frac{a}{b}$ belongs to $\FF_n \setminus \FF_{n-1}$ if and only if its continued fraction expansion
$\frac{a}{b}=[a_1,a_2,\dots, a_k]$ with $a_k>1$ is such that $\sum_{i=1}^k a_i = n$.

\noindent 
In this paper we shall study a family of \emph{signed generalized transfer operators} $\PP_q^{\pm}$ associated to
the map $F$, whose action on a function $f(x) : [0,1]\to \C$ is given by a weighted sum over the values of $f$ on the set
$F^{-1}(x)$, namely
\begin{equation} \label{ddeeff}
f(x) \longmapsto (\PP_q^{\pm} f)(x)=
\left(\frac{1}{x+1}\right)^{2q} \left[ f\left(
\frac{x}{x+1}\right) \pm f \left( \frac{1}{x+1} \right) \right]
\end{equation}
where $q$ is a real or complex parameter. The operator $\PP_1^{+}$ is referred to as the {\sl Perron-Frobenius} operator for
the map $F$: its fixed function is the density of an absolutely continuous $F$-invariant measure. In this case one easily
checks that the function $1/x$ has this property. However, since $1/x$  does not belong to $L^1([0,1],dx)$ the statistical
properties of the map $F$ have to be described in the framework of infinite ergodic theory \cite{Aa}. We
refer to
\cite{Bal} for a general review of transfer operator techniques in dynamical systems theory. 
Here, one motivation to study signed transfer operators arises from their appearing in  dynamical zeta functions such as
Selberg and Ruelle's (see \cite{DEIK}, Corollary 3.13, and also \cite{BI}).

\noindent
Using the Farey fractions, the iterates ${\PP_q^{\pm}}^n f$ of the
above operators can be expressed as suitable sums over the
\emph{Stern-Brocot tree}, the binary tree with root node $1$ and
whose $n$-th level $L_n$ is given by $L_n=\left(\FF_n \setminus
\FF_{n-1} \right) \cup S \left(\FF_n \setminus \FF_{n-1}\right)$,
where $S$ is the map $S: x\to 1/x$ and such that the elements of
$S\left(\FF_n \setminus \FF_{n-1}\right)$ are in reverse order. An important feature of
this tree is that each positive rational number appears as a vertex exactly once. The left part of the Stern-Brocot tree
(starting from the node $1\over 2$) is called the \emph{Farey tree}, with vertex-set $\Q \cap (0,1)$.

\begin{figure}[h]
\begin{center}
\includegraphics[width=8.0cm]{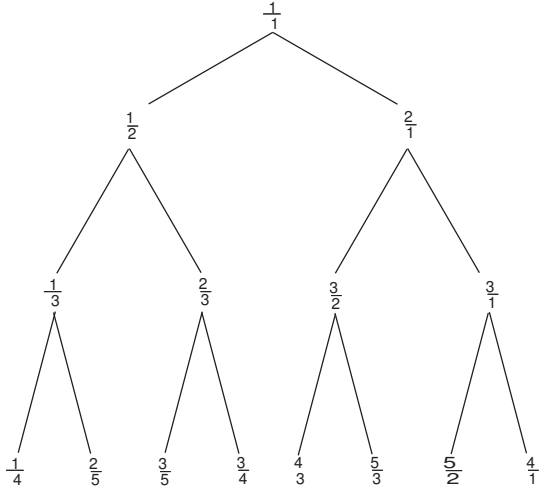}
\caption{{\small First four levels of the Stern-Brocot tree. }}
\label{indu}
\end{center}
\end{figure}

\noindent
An easy generalisation of Proposition 5.9 in \cite{DEIK} yields for all $x\in
\R_+$ and $q\in \C$,
\begin{equation}\label{SB}
({\PP_q^\pm}^n f) (x)=\sum_{\frac a b \in L_n} \frac{ f\left(
\frac{n_0(x, a/b)}{ax+b} \right) \pm f\left( \frac{n_1(x,
a/b)}{ax+b} \right)}{(ax+b)^{2q}}
\end{equation} where $n_0(x,a/b)=\mu x+\nu$
and $n_1(x,a/b)=(a-\mu)x+b-\nu$, for some $0\le \mu \le a$ and
$0\le \nu \le b$. In particular $n_0(x,a/b)+n_1(x,a/b)=ax+b$.

\noindent
In Section \ref{real-case} we prove 

\begin{theorem} \label{spettrone}
For each $q\in (0,\infty)$ there is a Hilbert space of analytic
functions $\HH_q$ on which the operators $\PP_q^{\pm}$ are
bounded, self-adjoint and iso-spectral. Their common spectrum is
given by $\set{0} \cup\, (0,1]$, with $(0,1]$ purely absolutely
continuous.
\end{theorem}
\noindent
\begin{remark}\label{leading}
From thermodynamic formalism it follows that  ${\cal P}_q^+$ for  $q\in (-\infty , 1)$, 
 when
acting on a suitable Banach space has a leading eigenvalue $\lambda
(q)\geq 1$  which is a differentiable and monotonically decreasing function
with 
$\lim_{q\to 1_-}\lambda (q)=1$ (and $\lambda (q)=1$ for all $q\geq 1$, see \cite{PS}). From the above theorem we see that
corresponding eigenfunction does not belong to the space ${\cal H}_q$ (for $q=1$ it is just the invariant density $1/x$).
Moreover
$$
\lambda (q) = \lim_{n\to \infty} {1\over n} \log ({\PP_q^+}^n 1) (0)\, .
$$
Note that by (\ref{SB}) we can write
$$
({\PP_q^+}^n 1) (0)= 2\, \sum_{{a\over b}\in {\cal F}_n\setminus \{{0\over 1}\} }b^{-2q} 
$$
and the above sum is equal to the {\sl partition function}
$Z_{n-1}(2q)$ at (inverse) temperature $2q$ of the number-theoretical
spin chain introduced by Andreas Knauf in \cite{Kn}. 
\end{remark}
\begin{remark} \label{cucu}
One easily checks that the function $f(x)=(1-x)/x$ is an eigenfunction of ${\cal P}_q^-$ for $q=1/2$ and eigenvalue $1$. But,
again, this function does not belong to ${\cal H}_{1\over 2}$.
\end{remark}
\noindent
There are interesting functional symmetries related to the eigenvalue equation for $\PP_q^{\pm}$, which can be rephrased in
terms of Hankel transforms. The construction of Section \ref{real-case} allows for a
complete account of the corresponding self-reciprocal functions in ${\rm
L}^2(\R_+)$, discussed in Section
\ref{selfrecip}. Finally, in Section \ref{poly} we characterise all polynomial
eigenvectors of
$\PP_q^{\pm}$ when
$q=-k/2$, $k\geq 0$. 

\section{The spectrum of
$\PP_q^{\pm}$ for real positive $q$} \label{real-case}

In this section we give the proof of Theorem \ref{spettrone},
hence in the sequel we restrict ourselves to the case $q\in
(0,\infty)$. The proof of the theorem follows from the results of
the following subsections.

\subsection{An invariant Hilbert space} \label{invariant-hilbert}

In this subsection we introduce a family of Hilbert spaces
$\HH_q$, where $q\in (0,\infty)$, and give the representation of
the operators $\PP_q^\pm$ on $\HH_q$. 

\begin{definition} \label{h0}
For $q\in (0,\infty)$ we denote by $\HH_q$ the Hilbert space of
all complex-valued functions $f$ which can be represented as a
generalised Borel transform
\begin{equation} \label{borel-def}
f(x)={\BB}_q [\varphi](x):= \frac{1}{x^{2q}}\ \int_0^\infty
e^{-\frac{t}{x}}\, e^t\, \varphi (t)\, m_q(dt) \qquad \varphi \in
{\rm L^2}(m_q)
\end{equation}
with inner product
\begin{equation}
(f_1,f_2) = \int_0^\infty \varphi_1(t)\, \overline{\varphi_2
(t)}\, m_q(dt) \quad \mbox{if }\ f_i=\BB_q [\varphi_i]
\end{equation}
and measure ($p=2q-1$)
\begin{equation}
m_q(dt)= t^p\, e^{-t}\, dt
\end{equation}
\end{definition}

\vsni 
\noindent
Function spaces related to that introduced above have been
used in \cite{Is}, \cite{GI} and \cite{Pre}. In \cite{Is} an
explicit connection between the approach presented here and
Mayer's work on the transfer operator for the Gauss map \cite{Ma}
is established by means of a suitable operator-valued power series.

\begin{remark} \label{borel-laplace}
For $q\in \C$, $\Re\ q >0$, the space $\HH_q$ can be regarded as a
complex Hilbert space. Setting
\begin{equation}
\chi_p(x):=x^p \qquad (p=2q-1)
\end{equation}
an alternative representation for $f\in \HH_q$ can be obtained by
a simple change of variable when $x$ is real and positive:
\begin{equation} \label{borel}
(\chi_p \cdot f)(x) = \int_0^\infty e^{-s}\, (\chi_p \cdot
\varphi)(sx)\, ds
\end{equation}
Note that a function $f \in \HH_q$ is analytic in the disk 
\begin{equation}\label{disk} D_1=\left\{x\in \C : \Re\, {\textstyle \frac{1}{x}} > {\textstyle
\frac{1}{2}}\right\}=\{x\in \C : |x-1|<1\}
\end{equation}
 In particular,
\begin{equation}
(\chi_{p} \cdot \varphi) (t) =\sum_{n=0}^\infty\ \frac{a_n}{n!}\,
t^n \quad \Longrightarrow \quad (\chi_{p}\cdot f)(x) =
\sum_{n=0}^\infty \ a_n\, x^n
\end{equation}
in the sense of formal power series. So the power series of
$\chi_{p}\cdot \varphi$ is obtained Borel transforming that of
$\chi_{p}\cdot f$, in the usual sense. This justify the name of
the integral transform (\ref{borel-def}).
\end{remark}
\begin{remark} \label{notin}
The invariant density $1/x$ for the Farey map, that is the fixed function of $\PP_1^+$, is the generalised
Borel transform (for $q=1$) of the function $\varphi(t) =1/t$ which, however, does not belong to ${\rm
L^2}(m_1)$.
\end{remark}
\noindent
Let us now study the Hilbert space ${\rm L^2}(m_q)$. First of all
we notice that the measure $m_q(dt)$ is finite, indeed
\begin{equation}\label{massa}
\int_0^\infty\ m_q(dt)=\Gamma (2q)
\end{equation}
Second, for the linearly independent family of functions
$f_n(t):={t^n\over n!}$ $(n\geq 0)$ we have
\begin{equation}
(f_n,f_m)=\frac{\Gamma(n+m+2q)}{n! m!}
\end{equation}
This implies that the (generalised) Laguerre polynomials
$L_n^{p}(t)$  ($n\ge 0$, $\Re\, p > -1$) by
\begin{equation}\label{laguerres}
e_n(t):=L_n^{p}(t) = \sum_{m=0}^n\ \comb{n+p}{n-m} \
\frac{(-t)^m}{m!}
\end{equation}
form a complete orthogonal basis in ${\rm L^2}(m_q)$, with
\begin{equation}\label{normaliz}
(e_n, e_m) = \frac{\Gamma(n+2q)}{n!}\ \delta_{n,m}
\end{equation}
Moreover, using (\cite{GR}, p.850) and (\ref{normaliz}) we get for
$m\le n$
\begin{eqnarray} \label{scalarprod}
(f_n,e_m) &=&(-1)^m \frac{\Gamma(n+2q)}{m!(n-m)!} = (-1)^m
\comb{n}{m}\ \| e_n \|^2 \nonumber \\
&=&(-1)^m \frac{\Gamma(n+2q)}{\Gamma(m+2q)\, (n-m)!} \| e_m \|^2
\nonumber \\
&=&(-1)^m \left( \frac{n+p}{n-m} \right) \| e_m \|^2
\end{eqnarray}
In particular $(f_n,e_n)=(-1)^n \| e_n \|^2$. Also note that
$(f_n,e_m)=0$ for $m>n$. Comparing to (\ref{laguerres}) we obtain
the following result
\begin{lemma}
For each  $n \in \N_0$ the numbers
$$a_{n,m}:=\left\{
\begin{array}{cl}
(-1)^m \comb{n+p}{n-m}  & {\rm if}\ m\le n \\[0.5cm]
0  & {\rm if}\ m>n
\end{array} \right.$$
are the Fourier coefficients of $f_n$ w.r.t the basis $(e_m)$,
i.e. $$a_{n,m}= \frac{(f_n,e_m)}{\| e_m\|^2}$$ Moreover
$$f_n = \sum_{m=0}^n\ a_{n,m} \, e_m \qquad
e_n = \sum_{m=0}^n\ a_{n,m} \, f_m$$
\end{lemma}

\begin{remark}
In particular, the $(n+1)\times (n+1)$ lower triangular  matrix
$A_n:=(a_{i,j})_{0\leq i,j \leq n} $ satisfies $A^2_n = I_{n+1}$.
Therefore, the operator $\Pi_n : {\rm L^2}(m_q)\to {\rm L^2}(m_q)$
acting as
$$\Pi_n : \sum_{s=0}^\infty c_s e_s \longrightarrow
\sum_{s=0}^\infty c_s\ \sum_{r=0}^n\ \frac{(f_r,e_s)}{\| e_s
\|^2}\ f_r =\sum_{r=0}^n\ d_r f_r$$ with
$$d_r:= \sum_{s=0}^r a_{r,s} c_s \quad \mbox{or} \quad
{\bf d}^{(n)} = A_n \, {\bf c}^{(n)}$$ where we have set ${\bf
c}^{(n)}=(c_0,c_1,\dots, c_n)^{T}$ and similarly for ${\bf
d}^{(n)}$, is the orthogonal projection onto the linear subspace
spanned by $(1,t,\frac{t^2}{2!}, \dots ,\frac{t^n}{n!})$.
\end{remark}

\noindent
Let us now consider the action of the transform $\BB_q$ on the
functions $(e_n)$ and $(f_m)$. We have
\begin{eqnarray}\label{unoo}
\BB_q [e_n](x) &=& \sum_{m=0}^n\ \Gamma(2q+m)\,
\comb{n+p}{n-m} \ \frac{(-x)^m}{m!} \nonumber \\[0.2cm]
&=&(n+1)_{p}\, (1-x)^n
\end{eqnarray}
where $(a)_p:=\Gamma (a+p)/\Gamma(a)=a(a+1)\cdots (a+p-1)$ is the
shifted factorial, and
\begin{equation} \label{duee}
\BB_q[f_n](x) = (n+1)_{p}\, x^n\, .
\end{equation}
The next result describes the action of $\PP_q^\pm$ on the Hilbert
space $\HH_q$.

\begin{proposition}\label{lilla}
For $q\in (0,\infty)$ the space $\HH_q$ is invariant for
$\PP_q^{\pm}$ and $\PP_q^\pm:\HH_q\to \HH_q$ are positive
operators, isomorphic to  self-adjoint compact perturbations of
the multiplication operator $M:{\rm L^2}(m_q)\to {\rm L^2}(m_q)$
given by
$$(M\varphi) (t)= e^{-t} \, \varphi(t)$$ More specifically
$$\PP_q^{\pm}\, \BB_q\, [\varphi] = \BB_q\, [P^\pm  \varphi]$$
where $P^\pm=M\pm N$ and $N: {\rm L^2}(m_q)\to {\rm L^2}(m_q)$ is
the symmetric integral operator given by
$$(N \varphi) (t) = \int_0^\infty \frac{J_p \left( 2 \sqrt{st}
\right)}{(st)^{p/2}}\ \varphi(s) \, m_q(ds)$$ where $J_p$ denotes
the Bessel function of order $p$.
\end{proposition}

\noindent \emph{Proof.} The representation of $\PP_q^{\pm}$ on
$\HH_q$ follows from a direct computation (see \cite{Is},
\cite{GI}). The positivity amounts to
\begin{equation} \label{positivity}
((M\pm N) \varphi,\varphi) \ge 0 \qquad \forall\ \varphi \in {\rm
L^2}(m_q),\quad \| \varphi \| =1
\end{equation}
and can be checked expanding $\varphi$ on the basis of
(normalised) Laguerre polynomials. Indeed, a calculation using
(\cite{GR}, pp.849-850) yields
$$\frac{(M e_n,e_n)}{\| e_n \|^2} =  2^{-2n-2q} \comb{2n+p}{n}$$ and
\begin{eqnarray}
\frac{(N e_n,e_n)}{\| e_n \|^2} &=& 2^{-n-2q} \comb{n+p}{n}
\ {}_2F_1(-n,n+2q;2q;1/2) \nonumber \\[0.2cm]
&=& 2^{-n-2q}  \ P_n^{(p,0)}(0)\nonumber
\end{eqnarray}
where $P_n^{(a,b)}(x)$ denotes the Jacobi polynomial (\cite{AAR},
p.99). Since
$$P_n^{(p,0)}(0) = (-2)^{-n} \, \sum_{k=0}^n (-1)^k \comb{n+p}{k}
\comb{n}{k}$$ and
$$\comb{2n+p}{n} = \sum_{k=0}^n \comb{n+p}{k}\ \comb{n}{k}$$ we get
$$
{((M\pm N) e_n,e_n)\over \Vert e_n \Vert^2}= {1\over 2^{2n+2q}} \sum_{k=0}^n (1\pm (-1)^{n-k}){ n+p \choose k}{n\choose
k} 
$$
and thus
(\ref{positivity}). Finally, $N\varphi$ can be written as
$\int_0^\infty k(s,t)\varphi (s) m_q(ds)$ with symmetric kernel
\begin{equation} \label{kernel}
k(s,t)= \frac{J_{p} \left( 2 \sqrt{st} \right)}{(st)^{p/2}}
\end{equation}
From the estimates $J_p(t) \sim 2^{-p}\,t^p/\Gamma (p+1)$ as $t\to 0^+$
and $J_p(t)= O(t^{-1/2})$ as $t\to \infty$ (\cite{E}, vol. II), we
see that the kernel $k(s,t)$ is bounded and continuous. \qed

\vsni
\noindent
 We can now describe the action of $P^\pm$ on $(e_n)$ and
$(f_n)$. Applying the integral representation (see \cite{E},
Vol. II, p.190)
$$ n!\,e^{-t}\, L_n^{p}(t) = \int_0^\infty \frac{J_{p} \left(
2 \sqrt{st} \right)}{(st)^{p/2}}\ s^n\, m_q(ds)$$ we get
\begin{equation} \label{emme-enne}
M^{-1} N f_n  = e_n \qquad M^{-1} N e_n=f_n
\end{equation}

\subsection{Functional symmetries}\label{funcequ}
Let introduce an isometry which turns out to be
useful for the characterisation of eigenfunctions of the operators
$\PP_q^\pm$. Let $\JJ_q$ be the involution defined by
\begin{equation} \label{funceq}
(\JJ_q f)(x) := \frac{1}{x^{2q}}\ f\left(\frac{1}{x}\right)
\end{equation}
and consider its action on the Hilbert space $\HH_q$. We have the
following

\begin{proposition} \label{IqProp}
For any $\varphi \in {\rm L^2}(m_q)$ it holds
\begin{equation}
\JJ_q\, \BB_q\, [\varphi] = \BB_q\, [J\, \varphi]
\end{equation}
where $J:= N\, M^{-1}$ is a bounded operator in ${\rm L^2}(m_q)$
with $\| J \| \leq 2\pi$. If moreover $\PP_q^\pm f = \lambda \,f$
for some $\lambda \ne 0$ then $f$ satisfies the functional equation
\begin{equation} \label{symmetri}
\JJ_q f = \pm f
\end{equation}
\end{proposition}

\noindent \emph{Proof.} The representation of $\JJ_q$ in $\HH_q$
is easily checked by first noting that for any $f \in \HH_q$ the
function $\JJ_q f$ can be written as an ordinary Laplace
transform, i.e.
\begin{equation} \label{lap}
f(x)= \BB_q [\varphi](x) \qquad \Longrightarrow  \qquad (\JJ_q
f)(x) = \int_0^\infty e^{-tx} (\chi_p \cdot \varphi) (t) dt
\end{equation}
and then using Tricomi's theorem (\cite{Sne}, p.165). Let us prove
the bound on $\| J \|$. Adapting formula (33) of \cite{RS}, vol.
IV, to our situation we get for all $\varphi \in {\rm L^2}(m_q)$
and $\lambda \in [0,1]$,
\begin{equation} \label{stima-rs}
\| N (M-\lambda)^{-1}\varphi \|^2 \le \int_0^1 \| N
(M-\lambda)^{-1}\varphi \|^2 d\lambda \le 2\pi
\int_{-\infty}^\infty \| N e^{i\tau M}\varphi \|^2  d\tau
\end{equation}
On the other hand we claim that
\begin{equation} \label{stima-rs-int}
\int_{-\infty}^\infty \| N e^{i\tau M}\varphi \|^2  d\tau \le 2\pi
\int_0^\infty e^{-t}\left(\int_0^\infty |J_p(2\sqrt{st})|^2
|\varphi (s)|^2 s^{p} e^{-s}ds\right) dt
\end{equation}
To prove (\ref{stima-rs-int}) we write
$$(N e^{i\tau M}\varphi)(t)= \int_0^\infty
\frac{J_p(2\sqrt{st})}{(st)^{p/2}}\ e^{i\tau e^{-s}}\,
\varphi(s)\, s^p\, e^{-s}ds$$ so that interchanging the order of
integration
$$\| N e^{i\tau M}\varphi \|^2 = \int_0^\infty |G(t,\tau)|^2 e^{-t}
dt$$ where we have set
\begin{eqnarray}
G(t,\tau) &=&  \int_0^\infty J_p(2\sqrt{st})\, e^{i\tau e^{-s}}
s^{p/2} \varphi(s)\, e^{-s}ds\nonumber \\
 &=& -\int_0^1 J_p(2\sqrt{-t \ln u})\, e^{i\tau u}(-\ln u)^{p/2}\,
 \varphi(-\ln u) du \nonumber
\end{eqnarray}
Equation (\ref{stima-rs-int}) now follows by applying
Fourier-Plancherel theorem:
\begin{eqnarray}
\int_{-\infty}^\infty |G(t,\tau)|^2 d\tau &=& 2\pi \int_0^1
|J_p(2\sqrt{-t \ln u})\, \varphi (-\ln u)|^2 (-\ln u)^{p} du
\nonumber \\
 &=& 2\pi \int_0^\infty |J_p(2\sqrt{st})|^2\, |\varphi
(s)|^2 s^{p} e^{-s} ds\nonumber
\end{eqnarray}
Hence, putting together (\ref{stima-rs}) and (\ref{stima-rs-int}),
we have
$$\| N (M-\lambda)^{-1}\varphi \|^2 \le 4\pi^2 \int_0^\infty
e^{-t} \left( \int_0^\infty |J_p(2\sqrt{st})|^2 |\varphi (s)|^2
s^{p} e^{-s} ds\right) dt$$ The right hand side is bounded above
by
$$4\pi^2 \| \varphi \|^2 \int_0^\infty e^{-t} \sup_{st\geq 0}
|J_p(2\sqrt{st})|^2 dt =: 4\pi^2 C \| \varphi \|^2$$ Using
$\sup_{x\geq 0}|J_p(2\sqrt{x})|^2=1$ we get $C=1$. Therefore
$$\| N (M-\lambda)^{-1} \|^2 \le 4\pi^2 \qquad \forall\
\lambda \in [0,1]$$ Choosing $\lambda =0$ we get $\| J \| \le
2\pi$ as claimed.

\noindent To finish the proof, we note that if $\varphi \in {\rm
L^2}(m_q)$ the functions $M\varphi$ and $N\varphi$ are bounded at
infinity. Therefore, if $f\in \HH_q$ satisfies $\PP_q^\pm f =
\lambda \,f$ with $\lambda \ne 0$, then $f$ extends analytically
from the disk $D_1$ to the half-plane $\set{ \Re\, x >0}$. In
addition the expression $(\PP_q^\pm f)(x)$ reproduces itself times $\pm 1$  if
transformed by substituting $1/x$ for $x$ and dividing through
$x^{2q}$. Hence (\ref{symmetri}) holds. \qed

\begin{remark}
Note that (\ref{funceq}) is only a necessary condition for $f$ to
be an eigenfunction (with $\lambda \ne 0$). For instance the
function $f(x)=x^{-q}$ (which does not belong to $\HH_q$) although
plainly satisfying (\ref{funceq}) for all $q\in (0,\infty)$ is an
eigenfunction of $\PP_q^{+}$ only for $q=1$ (with $\lambda =1$).
\end{remark}

\begin{remark}
Applying Proposition \ref{IqProp}, the eigenvalue equations
$\PP_q^{\pm}f=\lambda f$, with $\lambda \ne 0$, can be rewritten
as the three-term functional equations,
\begin{equation} \label{nuovooperatore}
\lambda\, f (x) - f(x+1) =\pm \frac{1}{x^{2q}} f \left( 1+
\frac{1}{x}\right)
\end{equation}
which for $\lambda=1$ are studied in \cite{Le} and \cite{LeZa}.
\end{remark}

\subsection{The spectrum of $P^\pm$ in ${\rm L^2}(m_q)$}
\label{sec-l2}

We are now reduced to study the spectrum of the operators $P^\pm$
in ${\rm L^2}(m_q)$. Let us start studying the operators
\begin{equation} \label{q-op}
Q^{\pm}=M^{-1}P^{\pm} = I\pm M^{-1} N
\end{equation}
We first show that they are bounded in ${\rm L^2}(m_q)$.

\begin{lemma}
We have $\| Q^\pm \| \le 1+2\pi$.
\end{lemma}

\noindent \emph{Proof.} The adjoint of the operator $J=NM^{-1}$
dealt with in the previous subsection exists and equals
$J^*=M^{-1} N$. A priori it is defined only on $D(M^{-1})$. Recall
however that $J^*$ is continuous if and only if $J$ is such and
$\| J^*\| = \| J \|$. The assertion now follows from Proposition
\ref{IqProp}. \qed

\vsni
\noindent
 Recall now the orthogonal basis of ${\rm L^2}(m_q)$
given by $e_n(t)$ (see (\ref{laguerres})) and the independent
family of functions $f_n(t)={t^n\over n!}$. We introduce the families of
functions
\begin{equation} \label{acca}
\ell_n^{\pm}(t):=e_n(t)\pm f_n(t), \ \qquad \ h_n^{\pm}(t) :=
e^{-t}( e_n(t) \pm f_n(t))
\end{equation}
and consider the linear manifolds spanned by them.

\begin{proposition} \label{invas}
The linear manifolds $\EE^{\pm} \subset {\rm L^2}(m_q)$ defined by
\begin{equation} \label{linman}
\EE^{\pm} := \set{ \sum_{n=0}^m c_n h_n^{\pm}\ :\ c_n \in \C,\
0\le n \le m, \ m\ge 0 }
\end{equation}
have the following properties:
\begin{enumerate}
\item they are fixed by the operators $\pm J$, i.e. $\pm J\varphi
= \varphi$, $\forall\, \varphi \in \EE^{\pm}$;

\item their intersection is the trivial subspace, i.e. $\EE^+ \cap
\EE^- =\set{0}$;

\item they are dense, i.e. $\overline{\EE^{\pm}} \equiv {\rm
Span}\, \set{h_n^{\pm}}_{n\geq 0} = {\rm L^2}(m_q)$.
\end{enumerate}
\end{proposition}

\noindent \emph{Proof.} We first use (\ref{unoo}) and (\ref{duee})
to get
\begin{equation} \label{tree}
\BB_q [h_n^{\pm}] (x) = (n+1)_{p}\ \frac{1\pm x^n}{(1+x)^{n+2q}}
\end{equation}
hence $\JJ_q \BB_q [h_n^{\pm}](x) = \pm \BB_q [h_n^{\pm}](x)$. Now
the first property follows upon application of Proposition
\ref{IqProp}.

\noindent
The second property follows at once from the fact that the
operator $J$ is an involution.

\noindent
Finally, from the proof of Proposition \ref{lilla} and
(\ref{emme-enne}) one readily gets that $(h_n^{\pm},e_n)>0$,
$\forall n \geq 0$. This yields the density of $\EE^{\pm}$ in
${\rm L^2}(m_q)$. \qed

\vsni
\noindent
 Let us now consider the functions $(\ell_n^\pm)$.
From the definition it follows that the function $\ell_n^+(t)$ is
a polynomial of degree $2k$ for $n=2k$ and $n=2k+1$, ($k\geq 0$);
whereas $\ell_n^{-}(t)$ has degree $2k+1$ for $n=2k+1$ and
$n=2k+2$, ($k\geq 0$). Moreover we have $(\ell_n^{\pm},e_n)=(1\pm
(-1)^n) \| e_n \|^2$ so that
\begin{equation} \label{ell-en}
\begin{array}{c}
(\ell_{2k+1}^{+},e_{2k+1}) = (\ell_{2k+2}^{-},e_{2k+2})=0 \\[0.2cm]
(\ell_{2k}^{+},e_{2k})=2 \| e_{2k} \|^2 \\[0.2cm]
(\ell_{2k+1}^{-},e_{2k+1})=2 \| e_{2k+1} \|^2
\end{array}
\end{equation}

\begin{proposition} \label{nucleo}
Let $H^\pm:={\rm Span}\, \set{ \ell_n^{\pm} }_{n\geq 0}$. Then
\begin{enumerate}
\item ${\rm L^2}(m_q) = H^+ \oplus H^- $;

\item $Q^{\pm}|_{H^{\pm}}=2\, I$ and $Q^{\pm}|_{H^{\mp}}=0$.
\end{enumerate}
\end{proposition}

\noindent \emph{Proof.}
\begin{enumerate}
\item By the relations (\ref{ell-en}), $H^+$ and $H^-$ do not have
non-zero common vectors, thus $H^+ \cap H^- = \set{0}$. Moreover,
let $\varphi \in {\rm L^2}(m_q)$ be such that $\varphi \, \bot
\,H^+ \oplus H^-$. Since $(\ell_n^{\pm},e_n)=(1\pm (-1)^n) \| e_n
\|^2$ we get $\varphi =0$.

\item We recall (\ref{emme-enne}),
$$M^{-1} N f_n  = e_n \qquad M^{-1} N e_n=f_n$$
From it we get
$$Q^{\pm}\, \ell_n^{\pm} = 2\, \ell^{\pm}_n \quad \mbox{and} \quad
Q^{\pm} \ell_n^{\mp}=0$$ For $\varphi = \sum_{n=0}^m c_n
\ell_n^{\pm}$ we have by linearity $Q^{\pm}\varphi = 2 \varphi$ so
that $\| Q^\pm \varphi \| = 2 \| \varphi \|$, independently of
$m$. This implies $Q^{\pm}\varphi = 2 \varphi$ for all $\varphi
\in H^{\pm}$. Hence $Q^{\pm}H^\pm \subseteq H^\pm$ and
$Q^{\pm}|_{H^{\pm}}=2\, I$. In the same way one proves that
$Q^{\pm}|_{H^{\mp}}=0$. \qed
\end{enumerate}

\begin{remark}
From the above it follows that the operators $Q^\pm$ are bounded
in ${\rm L^2}(m_q)$ with $\| Q^\pm \| =2$.
\end{remark}

\noindent
The operators $P^\pm$ are self-adjoint and positive on ${\rm
L^2}(m_q)$, hence the spectrum is real and positive. Moreover $\|
P^\pm \|\le \| Q \| \, \| M \| = 2$. Hence $\sigma(P^\pm)
\subseteq [0,2]$. From the previous results we have information on
the point spectrum $\sigma_p(P^\pm)$.

\begin{corollary} \label{spettrino}
In ${\rm L^2}(m_q)$ it holds ${\rm Ker}\, P^{\pm} = H^{\mp}$ and
$\sigma_{p}(P^\pm)=\{0\}$ with infinite multiplicity.
\end{corollary}

\noindent \emph{Proof.} We first observe that since ${\rm Ker}\,
M=\set{0}$ we have by Proposition \ref{nucleo} $${\rm Ker}\,P^\pm
={\rm Ker}\,(M Q^\pm)= {\rm Ker}\, Q^\pm = H^\mp$$ Now suppose
that $P^{\pm} \varphi = \lambda \varphi$ for some $0<\lambda \le
2$ and $\varphi \not\equiv 0$. Then $\varphi \in H^{\pm}$ and
hence $P^\pm\varphi = M Q^\pm \varphi = 2 M \varphi$. Therefore we
would have $(2M-\lambda) \varphi=0$ which implies $\varphi \equiv
0$. \qed

\vsni
\noindent
 To discuss the rest of the spectrum, we first
characterise in more detail the nature of the perturbation
operator $N$.

\begin{proposition} \label{openneq}
For $\Re \, q >0$ the operator $N: {\rm L^2}(m_q)\to {\rm L^2}(m_q)$ is nuclear (and
hence of the trace class). Its spectrum is given by
\begin{equation} \label{spettroN}
\sigma(N) = \set{0} \cup \set{ (-1)^{k} \alpha ^{2\, (q+
k)}}_{k\ge 0}
\end{equation}
where $\alpha = (\sqrt{5}-1)/2$ is the golden mean. Each
eigenvalue $\lambda_k \in \sigma(N)$ is simple and the
corresponding (normalised) eigenfunction $\psi_k$ is given by
\begin{equation} \label{eigenvectorr}
\psi_k(t)= \sqrt{ \frac{5^q\, k!}{\Gamma (k+2q)}}\
L_k^{p}(\sqrt{5}\, t) \,\exp{(-\alpha t)}
\end{equation}
\end{proposition}

\begin{corollary} \label{norma}
For $\Re\, q >0$ it holds
$${\rm tr} (N) = \frac{1}{\sqrt{5}} \ \alpha^{p} \quad
\mbox{and} \quad \| N \| = \alpha^{2\Re\, q}<1 $$
\end{corollary}

\noindent \emph{Proof of Proposition \ref{openneq}.} Expanding the
kernel of $N$ (see (\ref{kernel})) on the basis $(e_n)_{n\ge 0}$,
one get (see \cite{Sze}, p.102)
$$\frac{J_{p}\left( 2 \sqrt{st} \right)}{(st)^{p/2}} = \sum_{n=0}^\infty
e_n(s) \, \frac{e^{-t} \, t^n}{\Gamma (n+2q)}$$ This yields
$$N \varphi = \sum_{n\ge 0} (\varphi, e_n) \, g_n$$
where $g_n(t) = N e_n (t) = {e^{-t}t^n / n!}$. Since
$$\| e_n \| = \sqrt{\frac{\Gamma(n+2q)}{n!}} \qquad
\| g_n \| = \frac{\sqrt{\Gamma(2n+2q)}}{n!\, 3^{n+q}}$$ we have
$$\sum_n \| e_n \| \, \| g_n \| < \infty$$
and therefore $N$ is nuclear. To compute the spectrum of ${N}$ we
use the following Hankel transform (see \cite{E}, vol. II)
$$\int_0^\infty x^{p+\frac 1 2} e^{-bx^2} L_k^p(ax^2) J_p(xy) \sqrt{xy}\, dx =
\frac{(b-a)^ky^{p+\frac 1 2}}{2^{p+1}\, b^{p+k+1}}\
e^{-\frac{y^2}{4b}}\, L_k^p\left(\frac{ay^2}{4b(a-b)} \right)$$
which can be recast in terms of the operator $N$ as
$$N \left[ L_k^{p}(2at)\, e^{-(2b-1)t} \right] = \frac{(b-a)^k}
{2^{2q}\, b^{2q+k}}\ e^{-\frac{t}{2b}}\, L_k^{p} \left(
\frac{at}{2b(a-b)} \right)$$ This becomes an eigenvalue equation
in ${\rm L^2}(m_q)$ provided $2b=\alpha^{-1}$ and $2a=\sqrt{5}$.
The normalisation constant results from (\ref{normaliz}) noting
that
$$\| L_k^{p} (\sqrt{5}\, t) \,\exp{(-\alpha t)} \| =
\frac{1}{5^{\frac{q}{2}}} \ \| L_k^{p}(t) \|$$ This gives the
eigenfunctions $\psi_k$, and the proof is complete. \qed

\vsni
\noindent
 We now put together the previous results. We have
seen that for all $q\in (0,\infty)$ the operators $P^{\pm}= M\pm N$ when acting on ${\rm
L^2}(m_q)$ are self-adjoint and positive with $\| M \| =1$ and $\|
N \| = \alpha ^{2q}$.

\noindent
The operator $M$ is spectrally absolutely continuous (\cite{Ka},
p.520). Its spectrum, being the essential range of the multiplying
function, coincides with $[0,1]$. This means that in the
orthogonal decomposition ${\rm L^2}(m_q)={\rm H}_{ac}(M)\oplus
{\rm H}_s(M)$ of the Hilbert space into the subspace of absolute
continuity ${\rm H}_{ac}(M)=\Pi_{ac} (M) {\rm L^2}(m_q)$ and that
of singularity ${\rm H}_s(M)=\Pi_{s} (M) {\rm L^2}(m_q)$, we have
${\rm H}_s(M)=0$ (and thus $\Pi_{ac} (M)=I$).

\noindent
On the other hand $N_q$ is of the trace class. Therefore, applying
the Kato-Rosenblum theorem (see \cite{Ka}, p.542, or \cite{RS},
vol. III, p.26), it holds

\begin{proposition} \label{unitary}
The operator $M$ is unitarily equivalent to the spectrally
absolutely continuous part of $P^{\pm}$. Hence on ${\rm L^2}(m_q)$
we have $\sigma_{ac}(P^\pm) = (0,1]$.
\end{proposition}

\begin{remark}
Such equivalence is gained by means of the one-parameter family of
unitary operators
$$W(\tau) = e^{i\tau P} \, e^{-i\tau M} \qquad -\infty < \tau<
\infty$$ The (strong) limits $W_{\pm}$ of $W(\tau)$ as $\tau\to
\pm \infty$ are called \emph{wave operators} and $S=W_+^*W_-$ the
\emph{scattering operator}, which is unitary on ${\rm L^2}(m_q)$
to itself and commutes with $M$. The Kato-Rosenblum theorem says
that in this case the wave operators $W_\pm$ exist and are
complete, meaning that they are partial isometries with initial
domain ${\rm L^2}(m_q)$ and range ${\rm H}_{ac}(P)=\Pi_{ac} (P)
{\rm L^2}(m_q)$. Therefore we have $W_\pm^*W_\pm = I$, $W_\pm
W_\pm^*=\Pi_{ac} (P)$ and $P W_\pm= W_\pm M $ (see \cite{RS}, vol.
III, pp. 17-19).
\end{remark}

\noindent
Putting together Proposition \ref{lilla}, Corollary
\ref{spettrino} and Proposition \ref{unitary}, we get Theorem
\ref{spettrone}.
 
\section{Digression: self-reciprocal functions in ${\rm L^2}(\R_+)$}\label{selfrecip}

Given a continuous function $\phi$ on $\R_+$ and $q\in \C$, with $\Re \, q>0$ (or $\Re\, p > -1$),
the function $J\phi= NM^{-1}\phi$ considered in Section \ref{funcequ} can be viewed as a version of 
 its {\sl Hankel transform}, i.e. \be\label{hankel1} J\phi\, (t) :=
\int_0^\infty J_{p}(2\sqrt{st})\,\left({s\over
t}\right)^{p/ 2}\, \phi (s)\, ds \end{equation} 
We can also define the conjugate transform ${\tilde J}$ as 
\be
\qquad {\tilde J}:=\chi_q J \chi_{p}^{-1}
\end{equation}
or else 
\begin{equation}\label{hankel2}
{\tilde J} \phi\, (t) = \int_0^\infty
J_{p}(2\sqrt{st})\,\left({t\over s}\right)^{p/ 2}\, \phi
(s)\, ds \end{equation}

\noindent 
From the asymptotic estimates on $J_p(t)$ we see that the conditions on $\phi$ sufficient to give
the absolute convergence of the integral (\ref{hankel1}) are 
$\phi (t) =
O(t^ {-a})$ as $t\to 0^+$ and $\phi (t) =
O(t^ {-b})$ as $t\to \infty$ with $a  < 2\, \Re\, q$ and $b > \Re\, q+{1\over 4}$.
For the integral (\ref{hankel2}) we have the same conditions with $b > {5\over 4}-q$ and $a< 1$. 

\vsni
\noindent
Accordingly, the identity ${\cal J}_q f = \pm f$ for $f={\cal B}_q\,[\varphi]$ can be rephrased as a
{\sl self-reciprocity} property for the functions $\varphi$ and $\psi :=
\chi_p\cdot \varphi$, that is
\begin{equation}\label{equa} 
{\cal J}_q f
= \pm f \quad \Longrightarrow \quad  J \,\varphi=\pm \varphi\quad\hbox{and}\quad {\tilde J}
\psi=\pm\psi
\end{equation}
\begin{lemma}
If $\varphi \in  {\rm L^2}(\R_+)$ then $\varphi \in  {\rm L^2}(m_q)\cap {\rm L^2}(\R_+)$ provided
$\Re\, p\geq  0$.  Conversely, if $\varphi \in  {\rm L^2}(m_q)$ {\rm and} $J\varphi =\pm\, \varphi$ then $\varphi \in {\rm
L^2}(\R_+)$.
\end{lemma}
\noindent
{\sl Proof.} The first implication is immediate. The second follows from the asymptotic estimates on $J_p(t)$. $\qed$

\vsni
\noindent
Therefore, we shall study self-reciprocal functions in ${\rm L^2}(\R_+)$.
Moreover, by a change of 
variables the conditions (\ref{equa}) can be recast in the form that the function
\begin{equation} \label{variante}\qquad  \phi
(t)= 2^{-q+{1\over 2}}t^{p+{1\over 2}}\, \varphi \left({t^2\over 2}\right)=2^{q-{1\over 2}}t^{-p+{1\over 2}}\, \psi
\left({t^2\over 2}\right)  \end{equation} satisfies 
$K\,\phi\, = \pm \phi$
where $K$ is the symmetric version of the
Hankel transform given by 
\begin{equation} \label{kappa}
\qquad K \phi\, (t) := \int_0^\infty J_{p}(st)\,\sqrt{st}\, \phi
(s)\, ds. 
\end{equation}
For
${\Re}\, p >-1$ the simplest solution of $K\,\phi\, = \phi$ is $\phi(t)=\sqrt{2}\, t^{-{1\over
2}}$ which corresponds to $\varphi (t)= t^{-q}$ and
$\psi(t)=t^{\, q-1}$. This solution
has been already considered above and does not belong to
${\rm L^2}(\R_+)$. We refer to \cite{Tit}, Chap.9, for an analysis of the  equation $K\,\phi\, = \phi$ in ${\rm
L^2}(0,\infty)$.

\vsni
\noindent
For $a>0$, let 
$S_a : {\rm L^2}(\R_+) \to {\rm L^2}(\R_+)$ be given by
$(S_a \varphi) (t) := a^q \, \varphi (at)$. Then $J S_a = S_{1/a} J$. In particular, since $J\, e^{-t}=e^{-t}$
we have that $a^q e^{-at}$ and $a^{-q}e^{-t/a}$ is a Hankel transform pair for
all
$a>0$. Now, the
operator ${\tilde J}$ is an adjoint to $J$ in the sense
that $<\psi,J\varphi>= <{\tilde J}\psi, \varphi>$ with
$<\phi_1, \phi_2>:=\int_0^\infty\phi_1(t){\overline \phi_2}(t)dt$.
Whence, the
identity 
\begin{equation}\label{pair}
\qquad  \int_0^\infty a^{-q} e^{-t/a}\psi_1 (t)dt=\int_0^\infty a^q e^{-at}\psi_2 (t)dt 
\quad ,\quad a>0
\end{equation}
must hold whenever $\psi_1$ and $\psi_2$ is a pair w.r.t the  Hankel transform ${\tilde J}$.
If moreover ${\tilde \psi}_2$ is another Hankel transform of $\psi_1$ then 
 $\int_0^\infty e^{-at} (\psi_2 -{\tilde
\psi}_2) dt =0$ for all $a>0$ so that $\psi_2 ={\tilde
\psi}_2$ almost everywhere. Therefore the identity (\ref{pair}) is a necessary and sufficient
condition for $\psi_1$ and $\psi_2$ to be a pair w.r.t the  Hankel transform ${\tilde J}$.
Let moreover
\begin{equation}\label{mellin}
\qquad \qquad \psi^*(s) := \int_0^\infty \psi (t) \, t^{s-1}\, dt
\end{equation}
be the {\sl Mellin transform} of $\psi$. If there are two constants $a<b$ such that $\psi (t) =
O(t^ {-a})$ as $t\to 0^+$ and $\psi (t) =
O(t^ {-b})$ as $t\to \infty$ then the integral (\ref{mellin}) converges for $s$ in the strip $a<
{\Re}\, s <b$ and $\psi^*(s)$ is a holomorphic function in this strip.
\begin{remark}
If ${\PP_q}^{+}f =
\lambda \,f$ then one easily checks that
$$
\lambda=1+{f(1)\over f(0)}
\quad ,\quad {\lambda\over 2}(\lambda -1)={f(2)\over f(0)}
$$
Thus, if
$\lambda \ne 1$ we have $ f(0)\ne 0$ and  $$f(x) \sim f(0) \, x^{-2 q}, \quad
x\to \infty 
$$ Therefore, if $\Re \, q >0$ then the Mellin transform $f^*$ is analytic in the strip $0<\Re\, s <
2\,\Re\, q$ and in this region it holds
$$
({\cal J}_q f)(x)=f(x) \quad \Longrightarrow \quad f^*(s) = f^*(2q-s).$$
\end{remark}

\vsni
\noindent
Now, taking the Mellin transform of both sides in (\ref{pair}) we obtain
$$
\Gamma (1-s) \,\psi_1^*(s) = \Gamma (s+p)\, \psi_2^*(1-p-s)
$$
Note that if $\psi = \chi_p \cdot \varphi$,
then $\psi^*(s) = \varphi^*(s+p)$.
Moreover, Mellin transforming (\ref{variante}) gives
$$
\phi^*(s)= 2^{{s\over 2}-{3\over 4}} \varphi^*\left({\textstyle {s\over 2}+{p\over 2}+{1\over 4}} \right)=
2^{{s\over 2}-{3\over 4}} \psi^*\left({\textstyle {s\over 2}-{p\over 2}+{1\over 4}} \right) 
$$
Therefore, if we define the weighted transforms ${\bar \varphi}^*$, ${\tilde \psi}^*$ and ${\hat \phi}^*$ as
$$
 {\bar \varphi^*} (s) := {\varphi^*(s)\over \Gamma (s)},\quad  {\tilde \psi^*} (s) := {\psi^*(s)\over \Gamma (s+p)},\quad
{\hat \phi^*} (s) := 2^{{p\over 2}+1} { \phi^*(s) \over  \Gamma \left({\textstyle{s\over 2}+{p\over 2}+{1\over
4}}\right)}
$$
and taking into account that 
$1+p-\left({ {s\over 2}+{p\over 2}+{1\over 4}}\right) =  {1-s\over 2}+{p\over 2}+{1\over 4}$,
we have the following result:
\begin{proposition}\label{mellin}  The functions $\varphi, \psi, \phi \in {\rm L^2}(\R_+)$, related to each other by
(\ref{variante}), are jointly self-reciprocal, i.e.
$J \varphi = \pm \,\varphi$, ${\tilde J}\psi = \pm \,\psi$ and $K\phi =\pm \,\phi$, if and only if $${\bar
\varphi}^*(s)=\pm
\, {\bar \varphi}^*(1+p-s)\quad ,\quad {\tilde \psi}^*(s) =\pm\, {\tilde \psi}^*(1-p-s)\quad , \quad {\hat \phi}^*
(s) =
\pm
\, {\hat \phi}^*(1-s)$$
\end{proposition}

\vsni
\noindent
The sequences $h_n^\pm$ introduced in (\ref{acca}) were our first example of self-reciprocal
functions in ${\rm L^2}(\R_+)$, in that $Jh_n^\pm = \pm \, h_n^\pm$ for all $n\geq 0$. Even more interesting
self-reciprocal functions are provided by the conjugate sequences
$\varphi_n,
\psi_n\in {\rm L^2}(\R_+)$,
$n\geq 0$, defined for $\Re\, p>-1$ by
\begin{equation}\label{xipienne}
 \varphi_n(t) :=  \sqrt{{\textstyle{2^{p+1}\, n!\over \Gamma (n+p+1)}}}\, e^{-t} L_n^p(2t) , \quad   
\psi_n(t) :=(\chi_p \cdot  \varphi_n)(t) ,
\end{equation}
and satisfying the condition $<\varphi_n, \psi_m>=\, \delta_{n,m}$. They are related to the sequences $h_n^\pm$ by
(see \cite{E}, vol. II, p.192)
$$
\varphi_n= (-1)^n\,\sqrt{{\textstyle{2^{p+1}\, n!\over \Gamma (n+p+1)}}}\, \sum_{m=0}^n {n+p \choose n-m} (-2)^{m}
\left({h_{m}^++h_{m}^-\over 2}\right)
$$
Thus
$$
{J}\,\varphi_n =(-1)^n \sqrt{{\textstyle{2^{p+1}\, n!\over \Gamma (n+p+1)}}}\,\sum_{m=0}^n {n+p \choose n-m} (-2)^{m}
\left({h_{m}^+-h_{m}^-\over 2}\right)
$$
which, compared to (\ref{laguerres}), yields
\begin{equation}
\qquad {J}\,\varphi_n =(-1)^{n} \varphi_n \quad , \quad {\tilde J}\,\psi_n =(-1)^{n}\, \psi_n\, .
\end{equation}
Note that
\begin{equation}\label{tree1}
\qquad {\cal B}_q[\varphi_n](x) \;\,=\;\, (n+1)_{p}\, {{(1- x)^n\;\over\quad (1+x)^{n+2q}}}
\end{equation}
so that ${\cal J}_q {\cal B}_q[\varphi_n] =(-1)^n {\cal B}_q[\varphi_n]$, as expected (compare to (\ref{tree})).

\noindent
Moreover
\begin{equation} 
\qquad {\bar \varphi_n}^*(s) = {(p+1)_{n} \over n!}\, {}_2F_1(-n\, ,s\, ;p+1\, ;2)
\end{equation}
which satisfies the functional equation of Proposition \ref{mellin} 
because of Pfaff's identity (\cite{AAR}, Theorem 2.2.5) which implies
$${}_2F_1(-n\, ,b\, ;c\, ;2) = (-1)^n  \, {}_2F_1(-n\, ,c-b\, ;c\, ;2).$$
\vsni
\noindent
Finally, the orthonormal family $\{\phi_n\}$ of ${\rm L^2}(\R_+)$ given by
\begin{equation}
\qquad \phi_n(t):= \sqrt{{\textstyle{2\, n!\over \Gamma (n+p+1)}}}\, \, e^{-t^2/2}t^{p+{1\over 2}} 
L_n^p(t^2) 
\end{equation}
satisfies 
\begin{equation} \label{opera1}
\qquad K \phi_n = (-1)^n \phi_n, \quad n\geq 0.
\end{equation}
Thus, the families $\varphi_n, \psi_n, \phi_n$ furnish  {\sl a complete characterization of self-reciprocal functions in
${\rm L^2}(\R_+)$ for the Hankel transforms $J, {\tilde J}, K$}. 

\begin{remark}
The functions $\phi_n$ are also solutions of the
differential equation:
\begin{equation}
 \qquad \phi_n'' - \left({p^2-1/4\over t^2} +t^2 -4n-2p-2\right)\phi_n =0
\end{equation}
as one can check using, e.g., \cite{E}, vol. II, p.188.
More specifically, the second order differential operator $H$ given by\footnote{In quantum mechanics this corresponds to
the Schr\"odinger operator for a two-dimensional isotropic harmonic potential (see \cite{RS}, vol. II, p.161).}
\begin{equation}
\qquad H  := {1\over 2} \left( -{d^2 \over dt^2} + {p^2-1/4\over t^2} +t^2\right)
\end{equation}
has for real $p\geq 1$ a unique self-adjoint extension on ${\rm C_0^\infty}(\R_+)$ which has an
integer-spaced spectrum so that 
\begin{equation}\label{opera}
\qquad H\,  \phi_n = (2n + p+1) \,\phi_n, \quad n\geq 0.
\end{equation}
For $-1<p<1$ there is 
more than one self-adjoint extension, one of which, however, still satisfies (\ref{opera}). 
Comparing (\ref{opera1}) and (\ref{opera}) one may regard the unitary mapping $K$ of
${\rm L^2}(\R_+)$ onto itself as a hyperdifferential operator of the form ($2q=p+1$)
\begin{equation}
\qquad K  = e^{i\pi q} \exp{(-{i\pi\over 2} H)}
\end{equation}
and acting on a suitable class of analytic functions (see \cite{Bar} and \cite{Wo} for a discussion on this and related
correspondences).
\end{remark}
\section{Polynomial eigenfunctions of $\PP_q^\pm$ for $q=-{k/ 2}$.}\label{poly}
Although the eigenfunction
$f^{(q)}(x)$ corresponding to the leading eigenvalue $\lambda (q)$  does not belong to the space ${\cal H}_q$ (see Remark
\ref{leading}), we shall see that explicit expressions for
$\lambda (q)$ 	and $f^{(q)}(x)$ can be obtained when $q=-k/2$ with
$k$ a non-negative integer. Note that these values correspond exactly to the simple poles of $\Gamma (2q)$ and thus, by
(\ref{massa}), to the $q$-values where the measure $m_q$ has an infinite mass. On the other hand,
for $q=-k/2$ the operators ${\PP}_q^{\pm}$ take the form
$$
{\PP}_{-{k\over 2}}^{\pm}f (x )=(x+1)^k \left[
f\left( {x\over x+1}\right) \pm f\left( {1\over x+1}\right)\right] 
$$
so that they leave invariant the vector space $\oplus_{n=0}^k\C x^n$ of polynomials of degree  $ \leq k$. 
In particular we expect $f^{(-{k\over 2})}(x)$ is a polynomial of degree $k$ with real coefficients.

\vsni
\noindent
To warm up, a direct calculation yields
\begin{eqnarray}
&f^{(0)}(x) = 1,\qquad \lambda (0)=2,&\nonumber \\ \nonumber
&f^{(-{1\over 2})}(x) = x+1,\qquad \lambda (-{1\over 2})=3,&\\ \nonumber
&f^{(-1)}(x) = x^2+ {\sqrt{17}-1\over 2}\, x+1,\quad \lambda (-1)={5+\sqrt{17}\over 2},&\\
\nonumber 
&f^{(-{3\over2})}(x) = x^3+2x^2+2x+1,\qquad \lambda (-{3\over2})=7,&\\
\nonumber 
&f^{(-2)}(x) = x^4+{\sqrt{113}-1\over 4} \,x^3+3\, x^2+{\sqrt{113}-1\over 4}\,x+1,\quad
\lambda (-2)={11+\sqrt{113}\over 2}& \nonumber
\end{eqnarray}

\noindent
To say more we first need the following result.
\begin{lemma} \label{propp}
The $(k+1)\times (k+1)$ real positive matrix $M_k$ defined as

$$
M_k(i,j) := \left\{ \begin{array}{ll}
{k-i\choose j-i}  & \mbox{if } \ i<j\\[0.2cm]
\;2 & \mbox{if }\ i=j\\[0.2cm]
{i\choose j} & \mbox{if }\ i>j
\end{array} \right. \qquad (0\leq i,j\leq k)
$$
has the following properties:
\begin{enumerate}
\item the symmetry $M_k(i,j)=M_k(k-i,k-j)$ holds for all $0\leq i,j\leq k$;
\item the sum $S_i$ of the entries in row $i$ equals $S_i=2^i+2^{k-i}$;
\item the sum $R_j$ of the entries in column $j$ equals $R_j={k+2 \choose j+1}$;
\item if $M_k \Phi = \lambda \Phi$ with $\C^{k+1}\ni \Phi :=  (b_0, b_1, \cdots , b_{k})^T$ and $\lambda \ne
0$ then 
$\Phi$ is either a palindrome or  a skew-palindrome, i.e.  
$b_i=\pm \, b_{k-i}$ for $0\leq i\leq k$;
\item $\sigma (M_k) \subset \R$ for all $k\in \N \cup \{0\}$;
\item $1\in \sigma (M_k)$  for all $k\in \N$.
\end{enumerate}
\end{lemma}
\noindent
{\sl Proof.} 
1), 2) and 3) follow by direct computation. To prove 4) we write 
the eigenvalue equation componentwise:
$$
\lambda\,b_i= \sum_{j=0}^{k}M(i,j)\, b_j\qquad (0\leq i\leq k)
$$
which yields, using the symmetry 1), 
$$
\lambda\,b_{k-i}= \sum_{j=0}^{k}M(k-i,j)\, b_j=
 \sum_{j=0}^{k}M(k-i,k-j)\, b_{k-j} =\sum_{j=0}^{k}M(i,j)\, b_{k-j} 
$$
so that $M_k \Phi = \lambda \Phi$ if and only if $M_k \Phi' = \lambda \Phi'$ with $\Phi' :=  (b_k, b_{k-1},
\cdots , b_{0})^T$. 
If $\lambda$ is (geometrically) simple then $\Phi = \pm \Phi'$ because $\Vert \Phi\Vert=\Vert
\Phi'\Vert$ where $\Vert \;\; \Vert$ is the euclidean norm in $\C^{k+1}$. In other words,
$\Phi = \pm \Phi'$ is a necessary and sufficient condition for $\Phi$ and $\Phi'$ to be linearly dependent.
Assume now that $M_k \Phi = \lambda \Phi$ with $\lambda$ of geometric (and thus algebraic) multiplicity bigger than
$1$. If
$\Phi$ and
$\Phi'$ are linearly dependent then we are done. Suppose they are not. Then the vectors $\Psi_\pm:=\Phi \pm\Phi'$ should also
be two linearly independent eigenvectors. But this is impossible because $\Psi'_\pm=\pm \Psi_\pm$. This concludes the proof of
4).
 
\noindent
As for 5) we observe that $M_1$ is symmetric and for each $k\geq 2$ it is not difficult to realise that
one can construct recursively a positive symmetric
$(k+1)\times (k+1)$ matrix $N_k$ such that the product $M_k N_k$ is symmetric as well. For instance with $k=4$ one gets
$$M_4=\left(
  \begin{array}{ccccc}
  2&4&6&4&1\\
  1&2&3&3&1 \\
  1&2&2&2&1 \\ 
  1&3&3&2&1 \\
  1&4&6&4&2
  \end{array}\right)\qquad
N_4=\left(
  \begin{array}{ccccc}
  1&13&1&7&18\\
  13&1&3&2&13 \\
  1&3&3&7&1 \\ 
  7&2&7&1&7 \\
  18&13&1&7&1
  \end{array}\right)
$$  
Then apply Theorem 1 in \cite{DH}.
Finally, the vector $\Phi = (1,0,\cdots, 0,-1)^T$ always satisfies $M_k \Phi = \Phi$, which yields 6).
$\qed$
\begin{remark}
If one defines a pseudo-scalar product of $\Phi=(b_0, b_1, \cdots , b_{k})^T$ and
$\Psi=(c_0, c_1, \cdots , c_{k})^T$ as $<\Phi, \Psi>:=\sum_{i=0}^k b_i {\overline c_{k-i}}$ then
the symmetry stated in 1) amounts to $<M_k \Phi, \Psi> =
{ {<\Phi, M_k^T \Psi>}}$. Moreover, 4) implies that if 
$M_k \Phi = \lambda \Phi$ with $\lambda \ne 0$ then
$<\Phi, \Phi> = \pm \Vert \Phi \Vert^2$. 
\end{remark}
\begin{theorem} \label{spinnaker} Let $q=-k/2$, $k\geq 1$. The polynomial
\begin{equation}\label{autofu}
f(x)=\sum_{i=0}^{k}{k\choose i}\, b_i\,x^i
\end{equation}
satisfies ${\cal P}^\pm_q f=\lambda f$ with
$\lambda \ne 0$ if and only if the 
vector
$\Phi=(b_0, b_1, \cdots , b_{k})^T$
satisfies
$M_k \Phi = \lambda  \Phi$
and is either a palindrome (if ${\cal P}_q^+ f=\lambda f$) or a skew-palindrome (if ${\cal P}_q^- f=\lambda
f$).
\end{theorem}
\begin{corollary}  The eigenvector
corresponding to the simple positive maximal eigenvalue $\lambda
(-k/2)$ of $M_k$ is always palindromic and we have the bounds:
$$
s-1+h \leq \lambda \leq S-1+{1\over g}
$$
where  $$
S:=\max_i S_i = 2^{k}+1\quad , \quad s:=\min_i S_i =
\left\{
\begin{array}{ll}
2^{{k\over 2}+1}+2^{{k\over 2}-1}, & k\; \mbox{even }\\[0.3cm]
2^{{k+1\over 2}}+2^{{k-1\over 2}}, & k\;  \mbox{odd }
\end{array} \right.
$$
and
$$
h = {-s+2+\sqrt{s^2+4(S-s)} \over 2}\quad , \quad g= {S-2+\sqrt{S^2-4(S-s)} \over 2(s-1)}
$$
\end{corollary}
\noindent
{\sl Proof.} Put together the above and \cite{MM}, p.155, eq.(9). $\qed$

\vsni
\noindent
{\sl Proof of Theorem \ref{spinnaker}.} 
Setting
\begin{equation}\label{polyn}
f(x)=a_kx^k+a_{k-1}x^{k-1}+\cdots +a_{1}x+a_0
\end{equation}
the conditions ${\cal J}_q\, f=\pm f$ implies that the
sequence of coefficients $a_i$ is either a palindrome or a skew-palindrome, i.e.
$a_i=\pm \, a_{k-i}$, $(0\leq i\leq k)$.
Inserting the function $f(x)$ written above into (\ref{nuovooperatore}) with $q=-k/2$
and
$k\geq 1$ we get
\begin{eqnarray}
\lambda \, \sum_{i=0}^ka_i \, x^i
&=& \sum_{i=0}^k a_{i}\sum_{j=0}^{i}{ {i}\choose{j}}(x^j\pm x^{k-j})
 \nonumber\\
&=& \sum_{j=0}^k \left[\sum_{l=j}^{k}{ {l}\choose{j}} a_{l}\right](x^j\pm x^{k-j}) \\ \nonumber
&=& \sum_{i=0}^k \left[\sum_{l=i}^{k}{ {l}\choose{i}} a_{l}\pm \sum_{l=k-i}^{k}{ {l}\choose{k-i}}
a_{l}\right]\, x^i  \nonumber
\end{eqnarray}
which in both cases yields
\begin{equation}
\lambda\,a_i= \sum_{l=0}^{i}{ {k-l}\choose{k-i}} a_l + \sum_{l=i}^{k}{
{l}\choose{i}}a_l\qquad (0\leq i \leq k)
\end{equation}
Defining new coefficients $b_i$ so that
\begin{equation}
a_i={k \choose i}\, b_i\qquad (0\leq i \leq k)
\end{equation}
and using the identities
$$
{k-l \choose k-i}{k \choose l}={k \choose i}{i \choose l}\quad\hbox{and}\quad 
{l\choose i}{k \choose l}={k \choose i}{k-i \choose l-i}
$$
we see that the above recursion becomes
\begin{equation}\label{mainrecursion}
\lambda\,b_i= \sum_{l=0}^{i}{i\choose l} b_l + \sum_{l=i}^{k}{
{k-i}\choose{l-i}}b_l\qquad (0\leq i \leq k)
\end{equation}
and the proof is complete. $\qed$

\begin{example} 
For $k=4$ we find ${\rm sp}(M_4)=\left\{{11+\sqrt{113}\over 2},1,{11-\sqrt{113}\over
2},-1,-1\right\}$ and the corresponding
eigenvectors are
$$
\Phi_1= \left(\begin{array}{c} 1\\  {\sqrt{113}-1\over 16}\\ {1/ 2}\cr {\sqrt{113}-1\over 16}\\
1\end{array}\right)\quad
\Phi_2= \left(\begin{array}{c} 1\\ 0\cr 0\cr 0\\ -1\end{array}\right)\quad
\Phi_3= \left(\begin{array}{c} 1\\ {\sqrt{113}+1\over 16}\\ {1/ 2}\\ {\sqrt{113}+1\over 16}\\ 1\end{array}\right)$$
$$
\Phi_4= \left(\begin{array}{c} 3\\ 0\cr -2\\ 0\cr 3 \end{array}\right)\quad 
\Phi_5=\left(\begin{array}{c} 0\\ -1\\ 0\\ 1\\ 0\end{array}\right)
$$
Therefore the spectrum of $M_4$ yields three eigenfunctions for 
${\cal P}^+_{-2}$:
$$
h_1(x) = x^4+{\sqrt{113}-1\over 4} \,x^3+3\, x^2+{\sqrt{113}-1\over 4}\,x+1
$$
$$
h_3(x)= x^4+{\sqrt{113}+1\over 4} \,x^3+3\, x^2+{\sqrt{113}+1\over 4}\,x+1
$$
and
$$
h_4(x) = -3 x^4+12\, x^2-3 
$$
and two eigenfunction for ${\cal P}^-_{-2}$:
$$h_2(x)=x^4-1$$
and
$$
h_5(x) =4x(1-x^2).
$$
\end{example}

\begin{remark}\label{eigen1} Eigenvectors of $M_k$ to the eigenvalue $1$ are related to the period functions for
the modular group (see  \cite{CM}). In particular, for $k\in \N$ the eigenvectors
$(1,0,\cdots, 0,-1)^T$ correspond to the fixed functions $x^k-1$ of ${\cal P}^-_{-k/2}$ which yield the
even part of the period functions corresponding to  holomorphic Eisenstein forms of weight $k+2$. The odd parts are
computed below in Proposition \ref{eigen}. Other linearly independent (skew-palindromic and palindromic) eigenvectors with
eigenvalue
$1$  are expected for
$k\geq 10$, as they are related to (even and odd part of) holomorphic cusp forms \cite{A}.
\end{remark} 

\noindent
For the sake of completeness we end with the following result, a version of which is contained in \cite{CM}.
\begin{proposition}\label{eigen} Let $B_m$ denote the $m$-th Bernoulli number. For $k\in \N \cup \{0\}$ the
function $f_k(x) \in \oplus_{n=-1}^{k+1}\C x^n$ given by
$$
f_k(x) := {\zeta(-k)\over 2}
(1+x^{k})+(-1)^{k}k!{\sum_{-1\leq n\leq k+1}} {B_{n+1}\,
B_{k+1-n}\over (n+1)!(k+1-n)!}x^n
$$
satisfies ${\cal P}^+_{-k/2}f_k=f_k$ for $k$ even and
$f_k \equiv 0$ for $k$ odd.
\end{proposition}
\noindent
Two examples are
\begin{eqnarray}
f_0(x) &=& {1\over 12}\left[ x+{1\over x}-3\right]\nonumber \\
f_2(x) &=& {1\over 360}\left[5 x- \left(x^3+ {1\over
x}\right)\right]\nonumber
\end{eqnarray}
Note that for $k\geq 1$, the odd parts of the period functions mentioned in Remark \ref{eigen1} can be expressed as
${(-1)^{k}\over k!}(f_k(x)-{\zeta(-k)\over 2}
(1+x^{k}))$ \cite{Za1}.

\vsni
\noindent
{\sl Proof.} Consider the function $\psi_q (x)$
defined for $\Re \,q > 1$  by 
$$
\psi_q (x) =
{\zeta(2q)\over 2} (1+x^{-2q})+{\sum_{n,m\geq 1}} {(nx+m)^{-2q}}
$$ It is shown in \cite{Za2} that the function $\psi_q(x)$ has
an analytic extension into the complex $q$-plane with a simple pole
at $q=1$, and the analytic continuation satisfies (\ref{nuovooperatore}) with the sign $+$ and $\lambda =1$ for
all $q\in \C \setminus \{1\}$. Note that if $\Re \,q > 1$ then $\psi_q (\infty) =
{1\over 2} \, \zeta (2q)$.

\noindent
The proof then amounts to show that for $q=-k/2$ the
analytic extension of the function $\psi_q$
is precisely $f_k$. This is achieved using
standard Mellin transform techniques: start
from the identity
\begin{eqnarray}
{\sum_{n,m\geq 1}} {(nx+m)^{-2q}}&=&{1\over \Gamma (2q)}
\int_0^\infty {\sum_{n,m\geq 1}} e^{-t(nx+m)}
t^{2q-1}dt\nonumber \\
&=&{1\over \Gamma (2q)} \int_0^\infty {t^{2q-1}\over
(e^t-1)(e^{tx}-1)}   dt\nonumber
\end{eqnarray}
Recalling that
$$
{1\over e^t-1}=\sum_{r=-1}^\infty {B_{r+1}\over (r+1)!}\,
t^r={1\over t} -{1\over 2} +\sum_{l=1}^\infty {B_{2l}\over (2l)!}
t^{2l-1}
$$
we get
$$
{\sum_{n,m\geq 1}} {(nx+m)^{-2q}}={1\over \Gamma (2q)}
\sum_{k=-2}^\infty \int_0^\infty c_k(x) t^{k+2q-1}dt
$$
with $c_{-2}=1/x$, $c_{-1}=-{1\over 2}(1+1/x)$ and for $k\geq 0$
$$
c_k = {\sum_{-1\leq n\leq k+1}} {B_{n+1}\, B_{k+1-n}\over
(n+1)!(k+1-n)!}\, x^n
$$
Now $\Gamma (2q)$ has simple poles at $2q=-k$, with $k=0,1,2,\dots$, with
residues $(-1)^{k}/k!$
On the other hand the integral written above has simple poles at
$2q=-k$, with 
$k=-2,-1,0,1,\dots$ the residues being $c_{k}$.  Therefore the analytic
continuation of ${\sum_{n,m\geq 1}} {(nx+m)^{-2q}}$ has only two
simple poles at $q=1$ and $q=1/2$ and the claimed expression of
$f_k$ follows by taking the limit $2q\to -k$ with
$k\in \N \cup \{0\}$. The last assertion is a consequence of the identity $\zeta (-k)=-B_{k+1}/(k+1)$ for $k$ odd.
$\qed$

\end{document}